\documentclass[referee]{raa}            % referee version: for submission

\usepackage{graphicx,times}             %for PS/EPS graphics inclusion, new
\usepackage{natbib}
\usepackage{amssymb,amsmath}
\usepackage{url}
\usepackage{algorithmicx,algorithm,algpseudocode}

\begin{document}

   \title{The Richardson-Lucy Deconvolution method to Extract LAMOST 1D Spectra
%\,$^*$
%\footnotetext{$*$ Supported by the National Natural Science Foundation of China.}
}
%   \subtitle{I. Place Your Subtitle Here}

   \volnopage{Vol.0 (20xx) No.0, 000--000}      %%preserved for Editor. DOn't remove!
   \setcounter{page}{1}          %%starting page, preserved for Editor. DOn't remove!

   \author{Min Li
      \inst{1,2}
   \and Guangwei Li \thanks{corresponding auther}
      \inst{3}
   \and Ke Lv \thanks{corresponding auther}
      \inst{1} 
   \and Fuqing Duan
      \inst{4}
   \and Hasitieer Haerken
      \inst{4,2}
   \and Yongheng Zhao
      \inst{2}
   }
%% Here is an example of three authors come from different institutes.
%% For single author or all the authors from an institute, use "\inst{}" only

   \institute{University of Chinese Academy of Science, Beijing 100049, China; {\it luk@ucas.ac.cn}\\
%% Please give the E-mail address of the author, to whom future correspondence and
%% offprint requests will be sent.
        \and
             Key Lab for Optical Astronomy, National Astronomical Observatories, Chinese Academy of Sciences, Beijing 100101, China\\
        \and
             Key Laboratory of Space Astronomy and Technology, National Astronomical Observatories, Chinese Academy of Sciences, Beijing 100101, China; {\it lgw@bao.ac.cn}\\
        \and
             College of Information Science and Technology, Beijing Normal University, Beijing 100875, China\\
\vs\no
   {\small Received~~20xx month day; accepted~~20xx~~month day}}

\abstract{We use the Richardson-Lucy deconvolution algorithm to extract one dimensional (1D) spectra from LAMOST spectrum images. Compared with other deconvolution algorithms, this algorithm is much more fast. The practice on a real LAMOST image illustrates that the 1D resulting spectrum of this method has a higher SNR and resolution than those extracted by the LAMOST pipeline. Furthermore, our algorithm can effectively depress the ringings  that are often shown in the 1D resulting spectra of other deconvolution methods.	 
}

   \authorrunning{M.Li, G.-W.Li \& K.Lv }            %author_head in even pages
   \titlerunning{Deconvolution extraction based on RLI }  % title_head in odd pages

   \maketitle
%% The author head (on even pages) and the title head (on odd pages) will be
%% automatically extracted from \author{} and \title{}. Whenever the title is too long,
%% you will be asked to supply a shorter one by inserting either \authorrunning{} or
%% \titlerunning{} before \maketitle. Anyway, you can specify your own heads.
%%
%%
%% Note: In the following text body of your manuscript, please note several differences from
%%       other major journals:
%% (1) \subsection{Please Capitalize the First Letter of Each Notional Word in Subsection Title}
%% (2) Please Capitalize the First Letter of Each Notional Word in all tables' captions

%
%________________________________________________ sections below
%
\section{Introduction}           %% first-level sections will be auto-capitalized
\label{sect:intro}

Guoshoujing Telescope (also called Large Sky Area Multi-Object Fiber Spectroscopic Telescope, LAMOST)\citep{1996ApOpt..35.5155W, 2004ChJAA...4....1S,2012RAA....12..723Z, 2012RAA....12.1197C, 2012RAA....12.1243L} can obtains 4000 spectra in one exposure, which has collected more than 10 million spectra(see \url{http://dr7.lamost.org}). So far, about 400 peer reviewed papers based on LAMOST data are published, which help us more deeply understand our Galaxy. 
\par
The spectrum extraction method is a technique to convert a two-dimensional (2D) CCD image into  one-dimensional (1D) spectra, which can help astronomers explore the natures of celestial objects. The traditional extraction methods, including Aperture Extraction Method (AEM, hereafter), the Optimal Extraction Method and the Profile Fitting Method, are the most frequently used methods at present, which are discussed in detail in \citep{2019MNRAS.484.2403L}. Compared to the traditional methods, the deconvolution method is a completely different method, which tries to recover the 1D spectra by eliminating the instrumental profiles (Point Spread Function, PSF) on a 2D image. The deconvolution method is first presented  by \citet{2010PASP..122..248B}. In their work, a calibration matrix  is constructed from known instrumental profiles, and then inverted to calculate the resulting 1D spectra. As a result, the method can only extract a small piece of spectrum image, and cannot be used to extract a large spectrum image (for example, a LAMOST $4K \times 4K$ spectrum image) because of huge storage for the calibration matrix, let along calculation. Furthermore, the noise in a real image would destroy the resulting 1D spectra.
\par
To overcome the above shortcomings, \citet{2015PASP..127..552G} presented a deconvolution algorithm based on the Tikhonov regularization method (TDM, hearafter) for practical spectrum extraction. Firstly, they gave a method to obtain all PSFs which vary with positions on a CCD image. Secondly, the big spectrum image is divided into many small subimages, whose calibration matrixes can be easily stored and inverted during calculation. Thirdly, the Tikhonov regularization item can effectively suppress the noise. This algorithm is the first practical deconvolution method to extract 1D spectra from a multi-fiber spectrum image. The signal-to-noise ratio (SNR) and resolution of the resulting 1D spectra are both higher than traditional methods.
However, they did not discuss how to choose the best Tikhonov parameter. The choice of the parameter is a crucial problem, because a fixed Tikhonov parameter is not always appropriate for all fibers in a multi-fiber spectrum image.

In the last paper \citep{2019MNRAS.484.2403L}, we developed a deconvolution method based on adaptive Landweber iteration (ALI, hearafter), which can extract 1D spectra with the regularization parameter adaptively selected for every fiber. The SNR and resolution of 1D resulting spectra are both as high as those of the 1D resulting spectra extracted by TDM with a deliberately selected Tikhonov regularization parameter. %Besides, it can also eliminate the serious crosstalk.
\par
In this paper, the Richardson-Lucy Iteration deconvolution method is presented. The method can not only suppress the noise and improve both SNR and resolution of the resulting 1D spectrum, but also reduce the ringings in the 1D resulting spectrum. Besides, the algorithm can be easily programmed and runs fast. The Richardson-Lucy Iteration formula is given in Section~\ref{sect:Spectrum extraction}; Section~\ref{experiments and analysis} shows the extraction experiments on simulation 2D spectrum images;  In Section~\ref{sec:practical application}, we practice our method on a real LAMOST spectrum image; Finally, the conclusion is given in Section~\ref{sect:conclusion}.

%In 1996, we started a project to obtain Johnson $V$ and Str\"{o}mgren
%$uvby\beta$ photometry for the poorly studied variables of
%``pulsational interest''~$\cdots\cdots$.
%We used the three-channel high-speed photoelectric photometer designed for
%the Whole Earth Telescope campaign
%(\citealt{Nather+etal+1990, Jiang+Hu+1998}),
%and the four-channel Chevreton photoelectric photometer
%(\citealt{Michel+etal+1990, Michel+etal+1992}) dedicated to
%the STEPHI (STEllar Photometry International, \citealt{Michel+etal+1992}).

%% Authors can give a citation as 'Michel et al. 1992'.
%% You may also use \cite, \citep and \citet for citation, and use Table~1 or Figure~1
%% and so forth. Using \ref and \label for cross-references of Tables/Figures
%% is a good way in adjusting/adding/removing text, tables or figures.

\section{Spectrum extraction method based on Richardson-Lucy iteration}
\label{sect:Spectrum extraction}

Because the cross-talk on a LAMOST CCD is marginal, we only discuss how to extract 1D spectrum from an image with only one fiber. The image model can be found in \citep{2019MNRAS.484.2403L}, which is:

\begin{equation}
f(x,y)=\sum_{j=1}^{N}{g_j}{h_j(x-j,y)}+{\eta(x,y)}
\label{eq:discrete}
\end{equation}
where, $N$ is the number of rows, $f(x,y)$ and ${\eta(x,y)}$  are the count and the noise at the $x$th row and $y$th column on the CCD, respectively, ${g_j}$ is the flux at the $j$th row, ${h_j(x,y)}$ is the value of the PSF at the $j$th row on $(x, y)$, and $G = (g_1,g_2,...,g_N)$ is the 1D resulting spectrum. PSFs can be obtained by the method outlined in \citet{2015PASP..127..552G} and \citet{2019MNRAS.484.2403L}.
\par
The Richardson-Lucy Iteration algorithm \citep{1972JOSA...62...55R, 1974AJ.....79..745L} is a typical nonlinear iterative algorithm, which is widely applied in astronomical and medical image processing.

\par
We can assume that the count of each pixel is independent and obeys Poisson distribution. We denote $$a(x,y) = \sum_{j=1}^{N}{g_j}{h_j(x-j,y)}$$, which is the real spectrum image without noise. Then the likelihood function of the image is \citep{Zou2001}:
\begin{equation}
P(f(x,y)|G)=\prod_{x,y}\frac{a(x,y)^{f(x,y)}e^{-a(x,y)}}{f(x,y)!}
\label{eq:likelihood}
\end{equation}
Then, 
\begin{equation}
\ln P(f(x,y)|G)=\sum_{x,y}[f(x,y)\ln a(x,y) -a(x,y) -  \ln (f(x,y)!)]
\label{eq:likelihood_ln}
\end{equation}

Let $\frac{\partial}{\partial g_j}[ \ln P(f(x,y)|G) ] = 0, j=1, 2, 3, ..., N$. Then

\begin{equation}
\sum_{x,y}[\frac{f(x,y)h_j(x-j,y)}{a(x,y)} - h_j(x-j,y)]  =0
\label{eq:obj1}
\end{equation}
or

\begin{equation}
\sum_{x,y}[\frac{f(x,y)h_j(x-j,y)}{a(x,y)}] - 1  =0
\label{eq:obj2}
\end{equation}
where, $j = 1, 2, 3, ..., N$.
\par
To obtain the spectrum $G$, \citet{1986JOSAA...3.2121M} suggested the iteration formula:
\begin{equation}
g_j^{(k+1)} = g_j^{(k)} \Big\{\sum_{x,y}\frac{f(x,y)h_j(x-j,y)}{a(x,y)} \Big\}^p, j = 1,2,3, ..., N
\label{eq:obj2}
\end{equation}
where, $k$ is the number of iteration.
\par
If we set $p=1$, the above formula becomes Richardson-Lucy Iteration:
\begin{equation}
g_j^{(k+1)} = g_j^{(k)} \Big\{\sum_{x,y}\frac{f(x,y)h_j(x-j,y)}{a(x,y)} \Big\}, j = 1,2,3, ..., N
\label{eq:rl}
\end{equation}

\section{experiments on simulation images}
\label{experiments and analysis}
We use the similar construction method as \citep{2019MNRAS.484.2403L} to generate the simulation images. The simulation image is generated by an input 1D spectrum with 4,000 flux points given by the LAMOST pipeline convolved with PSFs with size of $13 \times 15$ pixels. These PSFs are the linear interpolations on the basic PSFs from emission lines on an arc image. At last, the poisson noise is added. The resulting image is $4,000 \times15$ pixels. 
\par
We extracted the 1D spectrum from the simulation image by Equation~\ref{eq:rl} for 10 iterations. Besides, we also performed TDM and ALI for comparison. The Tikhonov regulation parameter for TDM was set to $0.02$, which is the best value in this extraction. The block sizes in TDM and ALI were both set to 20, while the computation precisions were both 100.
\par
\begin{figure}
	\centering
	\includegraphics[width=\textwidth, angle=0]{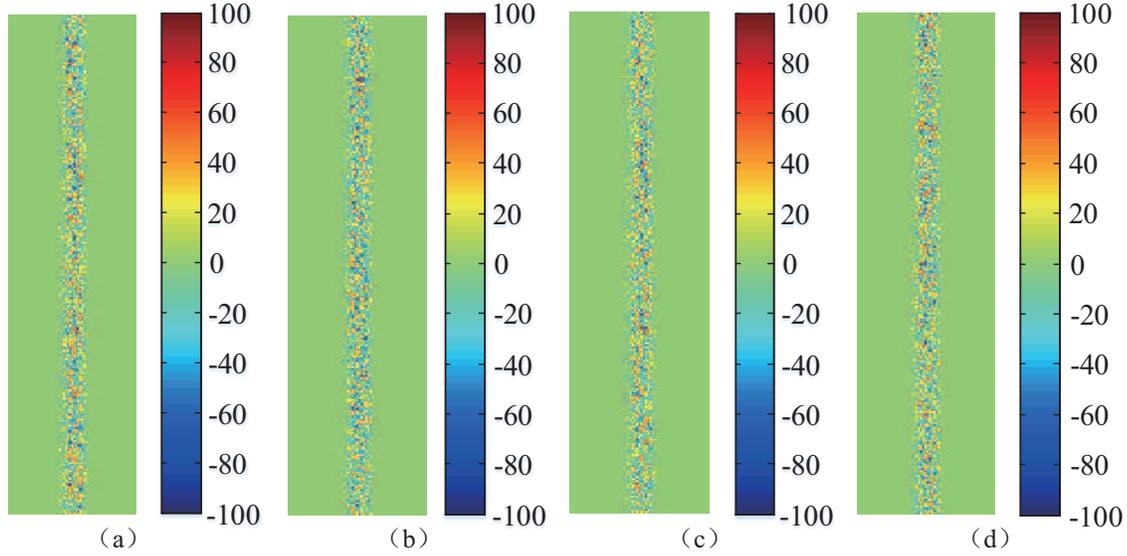}
	\caption{The original noise and 2D residuals of different methods. (a) the original Possion noise; (b), (c) and (d) are the 2D residuals of TDM, LIM and Equation ~\ref{eq:rl} for 10 iterations, respectively. }
	\label{two_res}
\end{figure}
\begin{figure}
	\centering
	\includegraphics[width=\textwidth, angle=0]{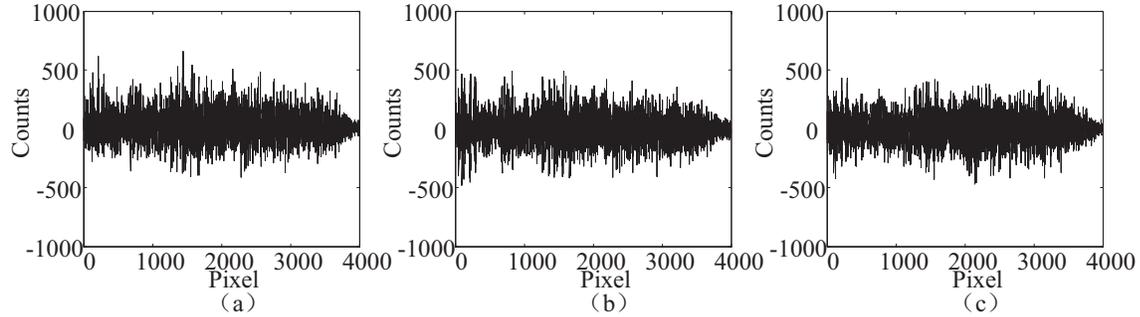}
	\caption{From left to right, panels show the 1D residuals of TDM, LIM and  Equation ~\ref{eq:rl} for 10 iterations, respectively.}
	\label{one_res}
\end{figure}

We ran these three algorithms on a DELL computer with CPU 3.30GHz and Operating system Windows 7 by the software MATLAB R2014a.
The original Poisson noise and the 2D residuals of different deconvolution methods are shown in Fig. \ref{two_res}, while 1D residuals is shown in Fig. \ref{one_res}. These two figures illustrate that their 2D residuals are Poisson noise, and their 1D residuals are all at the similar level. 
%{\color{red}We calculated the SNRs of different methods, which are shown in Table~\ref{Tab:snr}. The SNR is defined with the Equation 11 in our last paper \citep{2019MNRAS.484.2403L}. Table~\ref{Tab:snr} shows that the SNR of the spectrum extracted by our method is higher than others. }

\par
The the SNRs and computation times of different extraction methods are shown in Table~\ref{Tab:snr}, where the SNR is defined by the Equation 11 in \citep{2019MNRAS.484.2403L}. From Table~\ref{Tab:snr}, we can see that Richardson-Lucy Iteration algorithm is much faster than ALI and TDM, and the SNR of the resulting spectrum extracted by Richardson-Lucy Iteration algorithm is also higher than those extracted by ALI and TDM. TDM is the slowest, because it took too much time to inverse calibration matrixes.

\begin{table}
	\begin{center}
		\caption[]{SNRs and Computation Times of Different Methods}\label{Tab:snr}
		
		%%Please Capitalize the First Letter of Each Notional Word in table's caption
		
		\begin{tabular}{clcl}
			\hline\noalign{\smallskip}
			Methods &  SNR     & Time  (Seconds)           \\
			\hline\noalign{\smallskip}
			TDM  & 46.63     &   11.94                    \\
			LIM  & 47.08     &   6.05                      \\
	    	Equation ~\ref{eq:rl} for 10 iterations  & 50.03     &   1.64                  \\
	    	
			\noalign{\smallskip}\hline
		\end{tabular}
	\end{center}
\end{table}

\section{performance on a real LAMOST spectrum image}
\label{sec:practical application}
\begin{figure}
	\centering
	\includegraphics[width=\textwidth, angle=0]{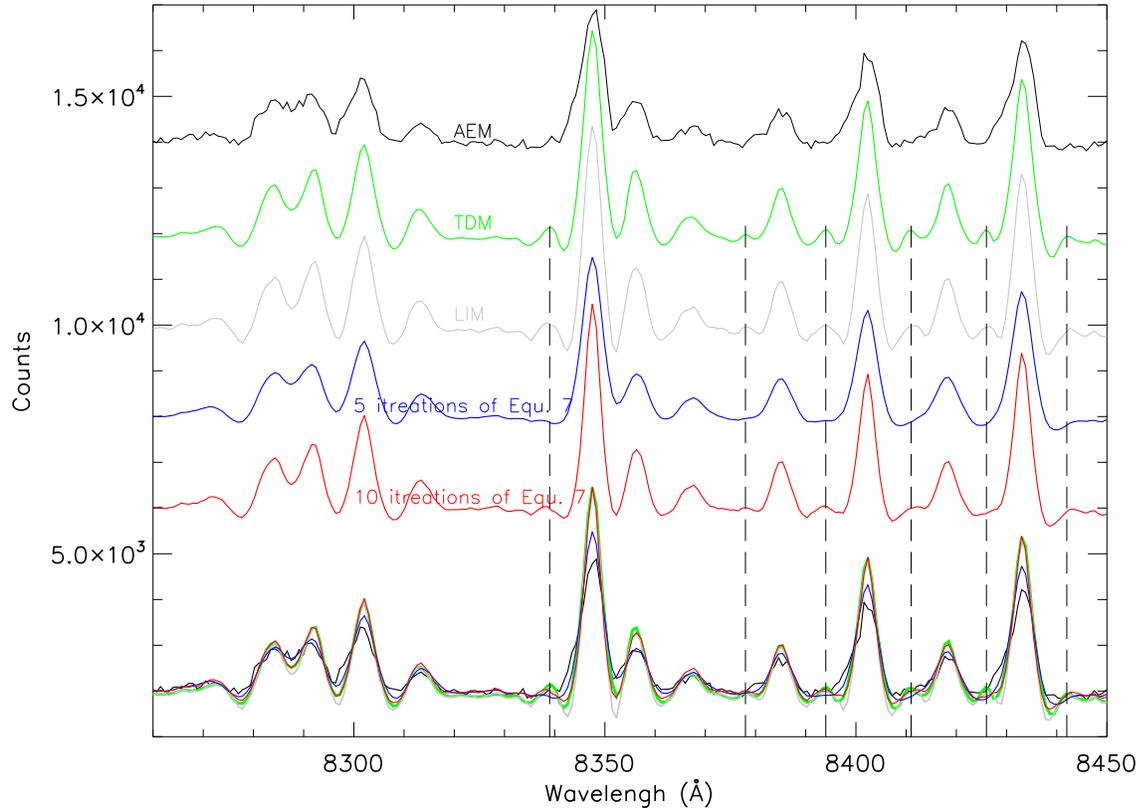}
	\caption{From top to bottom, the spectra are extracted by AEM, TDM, LIM, and Equation ~\ref{eq:rl} after five and 10 iterations, respectively. All these spectra are also overplotted at the bottom in the same colors.}
	\label{fig:real_extraction}
\end{figure}

We apply Equation~\ref{eq:rl} on a real LAMOST multi-fiber spectra image. A resulting 1D spectrum after five and 10 iterations are shown in blue and red, respectively in Figure ~\ref{fig:real_extraction}, while the spectra extracted by AEM, TMD and LIM are shown in black, green and grey, respectively. All spectra are overplotted together for direct comparison in the bottom of the figure.  As we all known, the Gibbs phenomenon, which also called ringing artifacts in signal processing, always occurs in the results of deconvolution methods. The dashed lines in  Figure~\ref{fig:real_extraction} indicate the positions of overshoots of ringings in the spectra of TMD and LIM.
\par
From the figure, we can see:\par
1) The spectra extracted by TMD, LIM and Equation ~\ref{eq:rl} all have higher SNRs and resolutions than that extracted by AEM. Furthermore, their emission lines are all more symmetric, which means that these three methods can correct the distortions of PSFs on CCD.
\par
2) The blue spectrum, which is the result of Equation ~\ref{eq:rl} after five iterations, even has no overshoots. Its resolution is lower than those of the spectra extracted by TMD and LIM, but higher than the spectrum extracted by AEM. 
\par
3) After 10 iterations of Equation ~\ref{eq:rl}, the resolution of the resulting spectrum become similar to those of the spectra extracted by TMD and LIM, but the amplitudes of ringings are much lower.

\section{Conclusions}
\label{sect:conclusion}
This paper describes a deconvolution extraction algorithm based on the Richardson-Lucy Iteration to extract 1D spectra from LAMOST spectrum images. Compared with the spectrum extracted by AEM, the resolution and SNR of the spectrum extracted by the Richardson-Lucy Iteration are both higher. Compared with spectra extracted by TDM and LIM, the ringings of the spectrum extracted by the Richardson-Lucy Iteration are much weaker. Furthermore, Richardson-Lucy Iteration is the fastest deconvolution method to extract 1D spectra from a LAMOST image.
   
\begin{acknowledgements}
This work is supported by the Joint Research Fund in Astronomy (U1531242)
under cooperative agreement between the National Natural
Science Foundation of China (NSFC) and Chinese Academy
of Sciences (CAS), the National Natural Science Foundation of China (No. 11673036) and the Interdiscipline Research Funds of Beijing Normal University.
\par	
Guoshoujing Telescope (the Large Sky Area Multi-Object Fiber Spectroscopic Telescope LAMOST) is a National Major Scientific Project built by the Chinese Academy of Sciences. Funding for the project has been provided by the National Development and Reform Commission. LAMOST is operated and managed by the National Astronomical Observatories, Chinese Academy of Sciences.

\end{acknowledgements}

\label{lastpage}

\end{document}